\documentclass{article}

\PassOptionsToPackage{numbers, compress}{natbib}

\usepackage[final]{neurips_2021_ml4ps}

\usepackage[utf8]{inputenc} 
\usepackage[T1]{fontenc}    
\usepackage{hyperref}       
\usepackage{url}            
\usepackage{booktabs}       
\usepackage{amsfonts}       
\usepackage{nicefrac}       
\usepackage{microtype}      
\usepackage{xcolor}         
\usepackage{amsmath}
\usepackage{graphicx}

\title{Automatic differentiation approach for reconstructing spectral functions with neural networks}

\author{
  Lingxiao Wang \\
    Frankfurt Institute for Advanced Studies\\
    \\
    Xidian-FIAS International Joint Research Center\\
    Frankfurt am Main,  D-60438, Germany \\
    \texttt{lwang@fias.uni-frankfurt.de}\\
   \And
  Shuzhe Shi \\
    Department of Physics\\
    McGill University\\
    Quebec H3A 2T8, Canada\\
    \\
    Center for Nuclear Theory\\
    Department of Physics and Astronomy\\
    Stony Brook University\\
    New York, 11784, USA\\
    \texttt{shuzhe.shi@stonybrook.edu}\\
  \AND
    Kai Zhou \\
    Frankfurt Institute for Advanced Studies\\
    Frankfurt am Main,  D-60438, Germany \\
    \texttt{zhou@fias.uni-frankfurt.de}
}
\begin{document}

\maketitle

\begin{abstract}
Reconstructing spectral functions from Euclidean Green’s functions is an important inverse problem in physics. The prior knowledge for specific physical systems routinely offers essential regularization schemes for solving the ill-posed problem approximately. Aiming at this point, we propose an automatic differentiation framework as a generic tool for the reconstruction from observable data. We represent the spectra by neural networks and set chi-square as loss function to optimize the parameters with backward automatic differentiation unsupervisedly. In the training process, there is no explicit physical prior embedding into neural networks except the positive-definite form. 
The reconstruction accuracy is assessed through Kullback–Leibler(KL) divergence and mean square error(MSE) at multiple noise levels. It should be noted that the automatic differential framework and the freedom of introducing regularization are inherent advantages of the present approach and may lead to improvements of solving inverse problem in the future. 
\end{abstract}

\section{Introduction}
The numerical solution to inverse problems is a vital area of research in many domains of science. In physics, especially quantum many-body physics, it’s necessary to perform an analytic continuation of function from finite observations which however is ill-posed~\cite{jarrell:1996bayesian,asakawa:2001maximum,tripolt:2019numerical}. It is encountered for example, in Euclidean Quantum Field Theory (QFT) when one aims at rebuilding spectral functions based on some discrete data points along the Euclidean axis. More specifically, the inverse problem occurs when we take a non-perturbative Monte Carlo simulations (e.g., lattice QCD) and try to bridge the correlator data points with physical spectra~\cite{asakawa:2001maximum,tripolt:2019numerical}. The knowledge of spectra will be further applied in transport process and non-equilibrium phenomena in heavy ion collisions~\cite{asakawa:2001maximum,rothkopf:2019bayesian}. Moreover, the inverse problem of rebuilding spectral function is not unique to strong interaction many-body systems, but have similar counterparts in quantum liquid and superconductivity~\cite{jarrell:1996bayesian}.

\paragraph{Related works} In past two decades, the most common approach in such reconstruction project is Bayesian inference which is a classical statistical learning method. It comprises the extra prior knowledge from the physical domain about the spectral function, so as to regularize the inversion task~\cite{asakawa:2001maximum,burnier:2013bayesian,Burnier:2014ssa}. 
For example, as two of axioms, the scale invariance and proper constant form prior are both embedded into the Bayesian approach with Shannon-Jaynes entropy, which is termed as maximum entropy method (MEM)~\cite{rothkopf:2019bayesian,Burnier:2014ssa}.
Besides, recent several studies have investigated reconstructing spectral functions through a neural network~\cite{kades:2020spectral,yoon:2018analytic,fournier:2020artificial,li:2020nett,zhou:2021application}. In a supervised learning framework, the prior knowledge is encoded in amounts of training data and the inverse transformation is explicitly unfolded through a training process~\cite{kades:2020spectral,yoon:2018analytic,fournier:2020artificial}. To alleviate the dependence of redundant training data, there are also
studies adopting the radial basis functions and Gaussian process~\cite{zhou:2021application,horak:2021reconstructing}.


\section{Problem statement}
\label{sec:sr}
Our inverse problem set-up is based on a Fredholm equation of the first kind, which takes the following form,
\begin{equation}
    g(t)=K \circ f:=\int_{a}^{b} K(t, s) f(s) d s,
\end{equation}
and the problem is to reconstruct the function $f(s)$ given the continuous kernel function $K(t,s)$ and the function $g(t)$ . In physical systems, $g(t)$ is always available in a discrete form. When dealing with a finite set of data points with finite uncertainty, the inverse transform becomes ill-conditioned or degenerated~\cite{fournier:2020artificial,caudrey:1982inverse}. In other words, the formal inversion is not a stable operation.

\paragraph{K\"allen--Lehmann(KL) spectral representation}
Henceforth in the paper, we discuss the uniqueness and accuracy of the spectral function by building the K\"allen--Lehmann(KL) representation~\cite{peskin:1995introduction},
\begin{align}
D(p) =\int_{0}^{\infty} K(p, \omega) \rho(\omega) d \omega \equiv \;& 
    \int_0^\infty 
    \frac{\omega \,\rho(\omega)}{\omega^2 + p^2} 
    \frac{\mathrm{d}\omega}{\pi}.
    \label{eq:corr}
\end{align}
Mock spectral functions are constructed using a superposed collection of Breit-Wigner peaks based on a parametrization obtained directly from one-loop perturbative quantum field theory~\cite{tripolt:2019numerical,kades:2020spectral}. Each individual Breit-Wigner spectral function is given by,
\begin{equation}
    \rho^{(\mathrm{BW})}(\omega)=\frac{4 A \Gamma \omega}{\left(M^{2}+\Gamma^{2}-\omega^{2}\right)^{2}+4 \Gamma^{2} \omega^{2}},
    \label{eq:bw}
\end{equation}
here $M$ denotes the mass of the corresponding state, $\Gamma$ is its width and $A$ amounts to a positive normalization constant. The multi-peak structure is built by combining different single peak modules together.

\section{Methods}
\label{sec:md}
In this section, we demonstrate vectorized formalism of our methodology which can be easily implemented by differential programming in Pytorch or other frameworks. For simplicity we take one $N_{p}$ point for $D(p)$ observation as example. One can directly extend it to multiple data points by making summation over them in calculating the gradients. 
\begin{figure}[htbp!]
  \centering
  \includegraphics[width = 13cm]{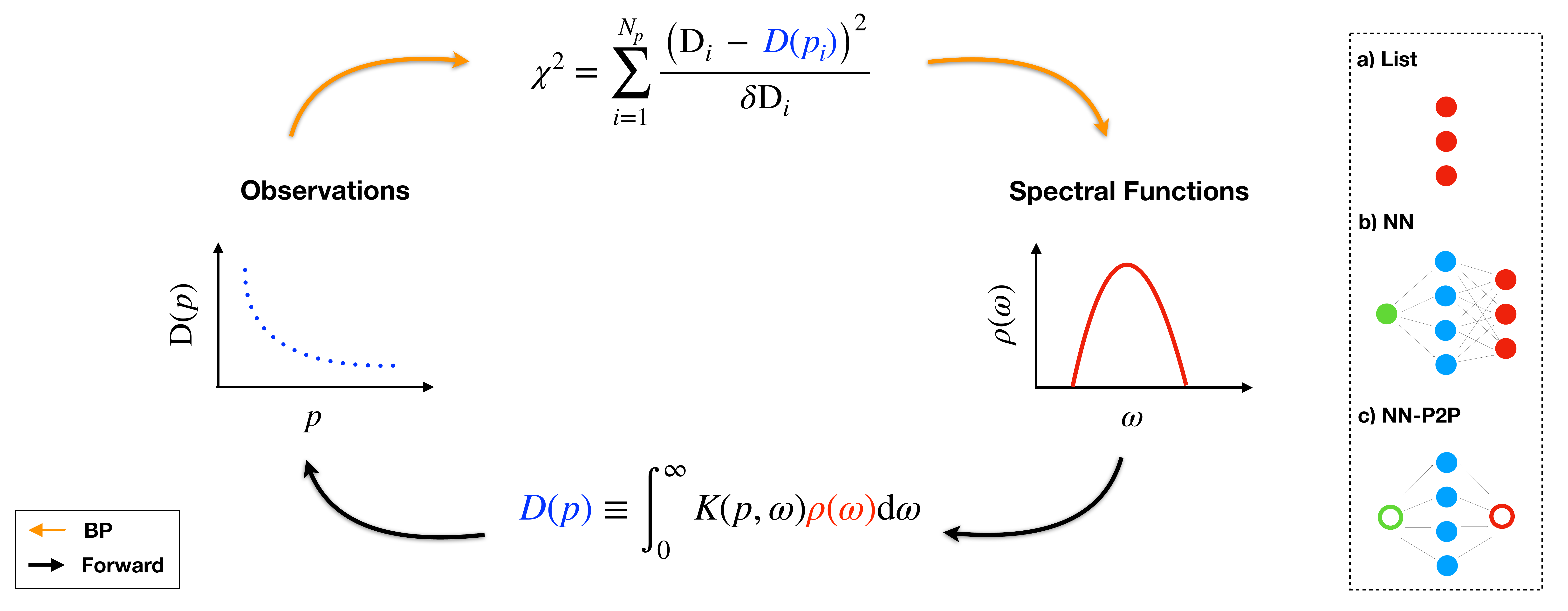}
  \caption{Automatic differential framework to reconstruct spectra from observations. Different architectures of representing spectrum with neural networks: a) \textbf{List}. $1$-layers neural network which is equivalent to the list of spectrum. b) \textbf{NN}. $L$-layers neural networks and the output is list of spectrum. c) \textbf{NN-P2P}. $L$-layers neural networks and the end-to-end nodes are $(\omega_i,\rho_i)$ pairwise. }
  \label{fig:frame}
\end{figure}
\paragraph{Spectral representations} we develop 3 architectures with different levels of non-local correlations among $\rho_{\omega_i}$ to represent the spectral functions with artificial neural networks(ANNs). The fist form is \textbf{List}, it is equivalent to set $L = 1$ without bias node, meanwhile, the differentiable variables are $\vec{\rho}=[\rho_1,\rho_2,\cdots,\rho_{N_{\omega}}]$ as Figure ~\ref{fig:frame} left panel a) shown. If one approximates the integration over frequencies $\omega_i$ to be summation over $N_\omega$ points at fixed frequency interval $\mathrm{d}\omega$, then it is suitable to the vectorized framework. The second representation is named as \textbf{NN}, in which we use $L$-layers neural network to represent the spectral function $\rho(\omega)$ with a constant input node $a^0=C$ and multiple output nodes $a^L = [\rho_1,\rho_2,\cdots,\rho_{N_{\omega}}]$. The width of the $l$-th layer is $n_l$, in which the correlation among discrete outputs is 
contained in a concealed form. The last way is to set input node as $a^0= \omega_i$ and single output node as $a^L = \rho_i$. It is termed as point-to-point neural networks (\textbf{NN-P2P}), in which the continuity of function $\rho(\omega)$ is a regularization defined in domains of input and output.

\paragraph{Automatic differentiation} The output of above representations is $\vec{\rho}=[\rho_1,\rho_2,\cdots,\rho_{N_{\omega}}]$, from which we can calculate the correlator as $D(p)=\sum_i^{N_{\omega}}\vec{\rho}\odot K(p,\vec{\omega})$ with $\vec{\omega}=[\omega_1,\omega_2,\cdots,\omega_{N_{\omega}}]$, where  `$\odot$' represents element-wise product. After the forward process, we can get both $\vec{\rho}$ and Loss $L=\chi^2 = \sum_i^{N_{\tau}} (\mathrm{D}_i - D(p_i))^2/D(p_i)$, where $\mathrm{D}_i$ is observed data at $p_i$ with $N_p$ points. To optimize the parameters of presentations $\{\theta\}$ with loss function, we implement the backward propagation (BP). The gradients for layer-$l$ is $\frac{\partial L}{\partial \theta^{[l]}}=\Delta^{[l]}$ and the input for backward propagation is,
\begin{align}
\Delta^{[L]}=\frac{\partial L}{\partial D({p})}K(p,\vec{\omega}).
\end{align}
With iteration loops in backward direction the gradients, $\Delta^{[l]}=\theta^{[l+1]\top}\Delta^{[l+1]}\odot\sigma'(Z^{[l]})$ can be used to optimize parameters $\{\theta\}$, where `$\top$' represents the transpose,  $\theta^{[l]}$ is weights matrix at layer-$l$, $Z^{[l]}$ is output of layer-$l$ and $\sigma(\cdot)$ is the corresponding activation function. 

\paragraph{Optimization strategy}
\label{para:opt}
The components of the framework are differentiable and therefore amenable to gradient descent. Due to the feasibility of regularizers in neural network representations, the optimization makes use of the Adam algorithm~\cite{kingma:2017adam} with weight decay\footnote{Weight decay is equivalent to $L^2$ regularization in stochastic gradient descent(SGD) when re-scaled by the learning rate~\cite{loshchilov:2019decoupled}.}. In training process, we obey an annealing strategy which is setting a tight regularization at beginning and loosen it repeatedly in fist 20000 epochs. The weight decay rate set as $10^{-4}$ and learning rate is $10^{-3}$ for all cases. The smoothness regularization contributed to loss function is written as $\lambda_s \sum_{i=1}^{N_{\omega}}(\rho_i- \rho_{i-1})^2$. Tight initial regularization is $\lambda_s =10^{-2}$. Besides the existing regularization of neural network itself, the only physical prior we enforce into the framework is the positive-definiteness of hadron spectral functions, which is introduced through using \textbf{Softplus} activation function at last layer as \textbf{Softplus}$(x) = log(1+e^x)$. It should be mentioned that the biases induced by using gradient descent-type optimizers are not avoided in our framework, but it 
could be improved by embedding ensemble learning strategies.

\newpage
\section{Numerical Results}
\begin{figure}[htpb!]
    \centering
    \includegraphics[width =12 cm]{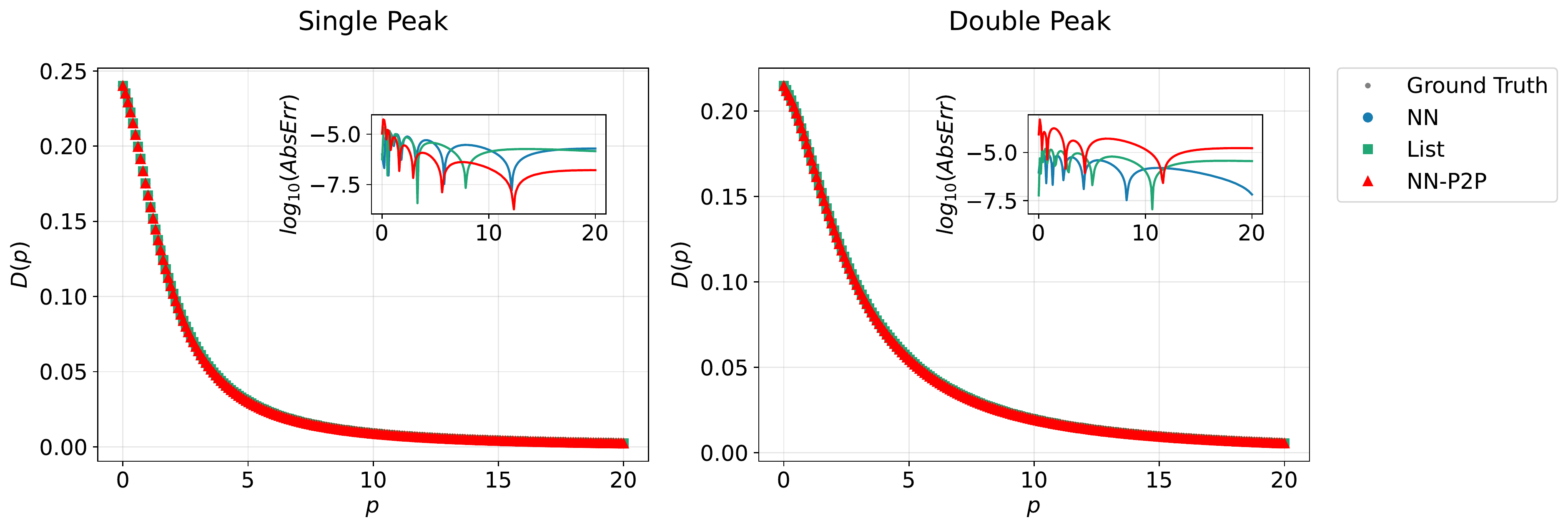}
    \caption{Performance of rebuilding correlation functions in two samples. The insert figures are showing absolute errors between observed correlation functions and the rebuilt.}
    \label{fig:per}
\end{figure}
In this section, to demonstrate the performance of our framework, we prepare two profiles of spectral functions from Eq.~\ref{eq:bw}. In Figure~\ref{fig:per}, the left correlation function is from a single peak spectrum with $A= 1, \Gamma = 0.3, M= 2.0$ and the right hand-side is from double peak profile with $A_1= 1, A_2 = 1.5, \Gamma_1 = 0.4, \Gamma_2 = 0.5, M_1= 2.5, M_2 =5.0$. Three representations are marked by green, blue and red dots, which are plot in high consistencies with observed correlators. Besides, to imitate the real-world observable data, we add Gaussian-type noise into mock data with $\Tilde{D}(p_i) = \mathrm{D}(p_i) + noise $ and $noise = \mathcal{N}(0,\epsilon)$. The reconstruction absolute error reaches $10^{-6}$ magnitude in all representations in the case with noise = $10^{-6}$. The corresponding rebuilt spectral functions are list in following Figure~\ref{fig:rec}. 
\begin{figure}[htpb!]
    \centering
    \includegraphics[scale=0.28]{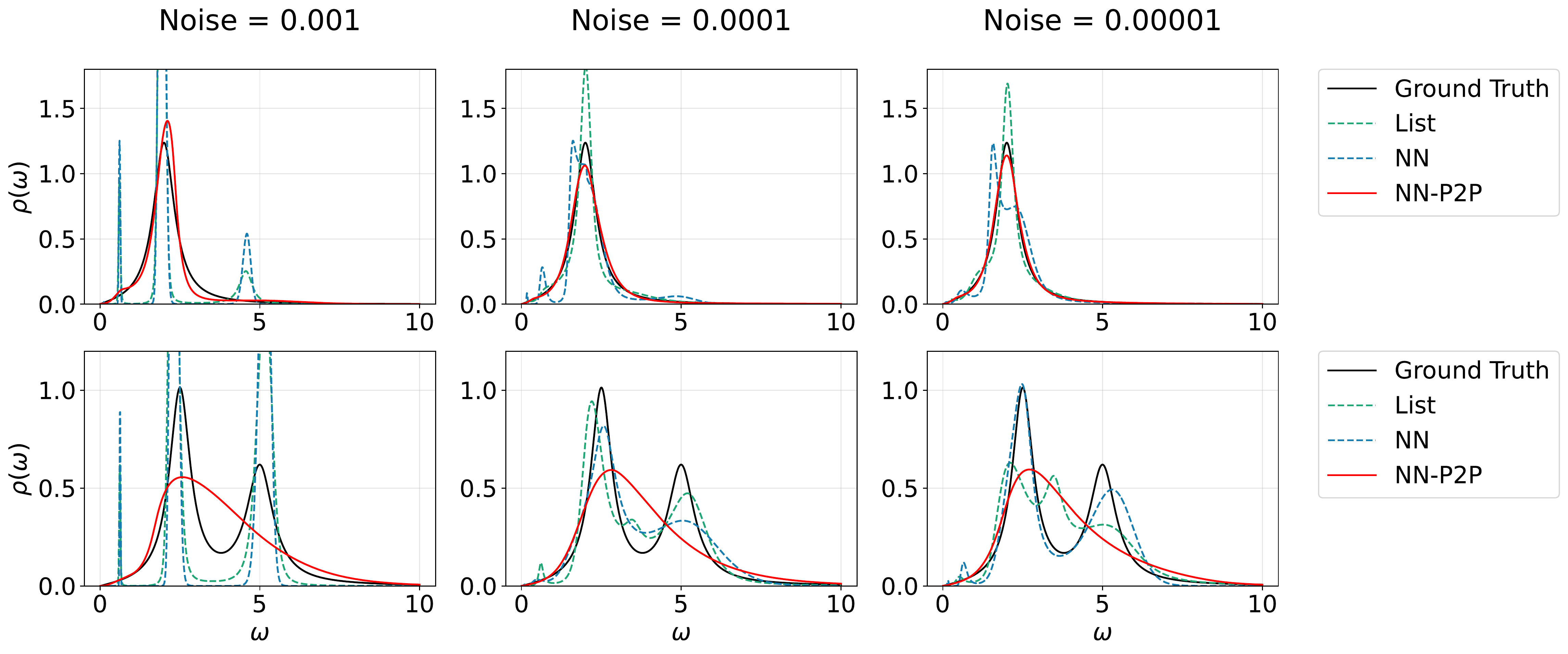}
    \caption{The predicted spectral functions from NN, List and NN-P2P. For left to right panels, different Gaussian noises are added to the correlation data with $\epsilon$ = 0.001, 0.0001 and 0.00001.}
    \label{fig:rec}
\end{figure}

Three representations are marked by green, blue and red lines in Figure~\ref{fig:rec}. They all show remarkable reconstruction performances for single peak case at noise level >0.0001. In which, \textbf{List} and \textbf{NN} behave oscillations around zero-point under different noise backgrounds. The rebuilding spectrum from \textbf{NN-P2P} do not oscillate even with noise smaller than 0.0001. This is especially important for such a task of extracting the transport coefficients from real-world lattice calculation data. Although the \textbf{List} representation has intense oscillations in double peak data, it successfully unfold the two peaks information from correlators even with noise $\epsilon$ = 0.001.
\begin{figure}[htpb!]
    \centering
    \includegraphics[width =12 cm]{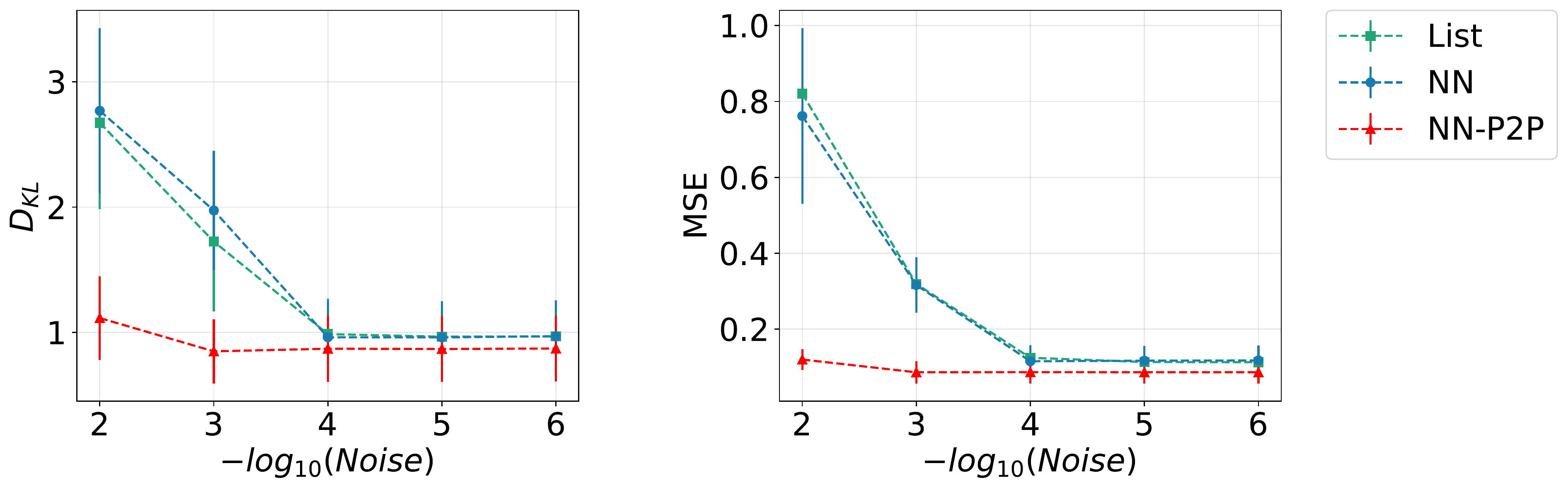}
    \caption{Evaluation to the reconstruction with KL divergence and MSE on mock data sets. Red triangle marks the NN-P2P, blue circles are NN representation and green cube labels the List.}
\end{figure}

To assess the reconstruction performance quantitatively, we introduce MSE and Kullback–Leibler(KL) divergence for rebuilt spectra $q$ and the ground truth $p$ as $D_{\mathrm{KL}}(P \| Q)=\int_{0}^{\infty} p(\omega) \log \left({p(\omega)}/{q(\omega)}\right) d \omega$. At multi-magnitude noises, the NN-P2P keeps consistent performances compared with the other two representations. Although it misses the second peak which may appear in the case of bimodal, the calculations of different order momentum from spectral function will not be disturbed. Intuitively speaking, in NN-P2P representation, there are series of 1-order differentiable modules 
between input $\omega$ node and output $\rho$ node, in which the continuity of function $\rho(\omega)$ is naturally preserved. It makes the NN-P2P has a better performance in single peak case but failed in reconstructing double peaks which needs a looser restriction.

\section{Conclusions}
We present an automatic differentiation framework as a generic tool for unfolding spectral functions from observable data. The representations of spectra are used with 3 different neural network architectures, in which the modern optimization algorithm will be naturally employed. Although the inverse problems cannot be fully-solved in our framework, the remarkable performances of reconstructing spectral functions suggest that the framework and the freedom of introducing regularization are inherent advantages of the present approach. In recent progress, we compare our approaches with the maximum entropy method (MEM) and discuss their merits and drawbacks~\cite{wang:2021reconstructing}. In future works, we will explore more neural network representations and a potential direction is to design specific neural networks with physics rules. This new paradigm provides us with a practical toolbox, in which solving inverse problem transfers into designing proper representations of solutions. It may lead to improvements in solving actual problems in e.g. optics, material design, and medical imaging.

\begin{ack}
We thank Heng-Tong Ding, Swagato Mukherjee and Gergely Endr\"odi for helpful discussions.  The work is supported by (i) the BMBF under the ErUM-Data project (K. Z. and L. W.), (ii) the AI grant of SAMSON AG, Frankfurt (K. Z.), (iii) Natural Sciences and Engineering Research Council of Canada (S. S.), (iv) the Bourses d'excellence pour \'etudiants \'etrangers (PBEEE) from Le Fonds de Recherche du Qu\'ebec - Nature et technologies (FRQNT) (S. S.), (v) U.S. Department of Energy, Office of Science, Office of Nuclear Physics, grant No. DE-FG88ER40388. (S. S.). K. Z. also thanks the donation of NVIDIA GPUs from NVIDIA.
\end{ack}

\bibliographystyle{unsrt}
\bibliography{Spectrum.bib}

\begin{thebibliography}{10}

\bibitem{jarrell:1996bayesian}
Mark Jarrell and J.~E. Gubernatis.
\newblock Bayesian inference and the analytic continuation of imaginary-time
  quantum {{Monte Carlo}} data.
\newblock {\em Physics Reports}, 269(3):133--195, May 1996.

\bibitem{asakawa:2001maximum}
M.~Asakawa, Y.~Nakahara, and T.~Hatsuda.
\newblock Maximum entropy analysis of the spectral functions in lattice
  {{QCD}}.
\newblock {\em Progress in Particle and Nuclear Physics}, 46(2):459--508,
  January 2001.

\bibitem{tripolt:2019numerical}
Ralf-Arno Tripolt, Philipp Gubler, Maksim Ulybyshev, and Lorenz {von Smekal}.
\newblock Numerical analytic continuation of {{Euclidean}} data.
\newblock {\em Computer Physics Communications}, 237:129--142, April 2019.

\bibitem{rothkopf:2019bayesian}
Alexander Rothkopf.
\newblock Bayesian techniques and applications to {{QCD}}.
\newblock {\em arXiv:1903.02293}, March 2019.

\bibitem{burnier:2013bayesian}
Yannis Burnier and Alexander Rothkopf.
\newblock Bayesian {{Approach}} to {{Spectral Function Reconstruction}} for
  {{Euclidean Quantum Field Theories}}.
\newblock {\em Phys. Rev. Lett.}, 111(18):182003, October 2013.

\bibitem{Burnier:2014ssa}
Yannis Burnier, Olaf Kaczmarek, and Alexander Rothkopf.
\newblock Static quark-antiquark potential in the quark-gluon plasma from
  lattice {{QCD}}.
\newblock {\em Phys. Rev. Lett.}, 114(BI-TP-2014-21):082001, February 2015.

\bibitem{kades:2020spectral}
Lukas Kades, Jan~M. Pawlowski, Alexander Rothkopf, Manuel Scherzer, Julian~M.
  Urban, Sebastian~J. Wetzel, Nicolas Wink, and Felix P.~G. Ziegler.
\newblock Spectral reconstruction with deep neural networks.
\newblock {\em Phys. Rev. D}, 102(9):096001, November 2020.

\bibitem{yoon:2018analytic}
Hongkee Yoon, Jae-Hoon Sim, and Myung~Joon Han.
\newblock Analytic continuation via domain knowledge free machine learning.
\newblock {\em Phys. Rev. B}, 98(24):245101, December 2018.

\bibitem{fournier:2020artificial}
Romain Fournier, Lei Wang, Oleg~V. Yazyev, and QuanSheng Wu.
\newblock Artificial {{Neural Network Approach}} to the {{Analytic Continuation
  Problem}}.
\newblock {\em Phys. Rev. Lett.}, 124(5):056401, February 2020.

\bibitem{li:2020nett}
Housen Li, Johannes Schwab, Stephan Antholzer, and Markus Haltmeier.
\newblock {{NETT}}: Solving inverse problems with deep neural networks.
\newblock {\em Inverse Problems}, 36(6):065005, June 2020.

\bibitem{zhou:2021application}
Meng Zhou, Fei Gao, Jingyi Chao, Yu-Xin Liu, and Huichao Song.
\newblock Application of radial basis functions neutral networks in spectral
  functions.
\newblock {\em arXiv:2106.08168}, June 2021.

\bibitem{horak:2021reconstructing}
Jan Horak, Jan~M. Pawlowski, Jos{\'e} {Rodr{\'i}guez-Quintero}, Jonas Turnwald,
  Julian~M. Urban, Nicolas Wink, and Savvas Zafeiropoulos.
\newblock Reconstructing {{QCD Spectral Functions}} with {{Gaussian
  Processes}}.
\newblock {\em arXiv:2107.13464}, July 2021.

\bibitem{caudrey:1982inverse}
P.~J. Caudrey.
\newblock The inverse problem for a general {{N}} \texttimes{} {{N}} spectral
  equation.
\newblock {\em Physica D: Nonlinear Phenomena}, 6(1):51--66, October 1982.

\bibitem{peskin:1995introduction}
Michael~E. Peskin and Dan~V. Schroeder.
\newblock {\em An {{Introduction To Quantum Field Theory}}}.
\newblock {Westview Press}, {Reading, Mass}, first edition edition edition,
  October 1995.

\bibitem{kingma:2017adam}
Diederik~P. Kingma and Jimmy Ba.
\newblock Adam: A {{Method}} for {{Stochastic Optimization}}.
\newblock {\em arXiv:1412.6980}, January 2017.

\bibitem{loshchilov:2019decoupled}
Ilya Loshchilov and Frank Hutter.
\newblock Decoupled {{Weight Decay Regularization}}.
\newblock {\em arXiv:1711.05101}, January 2019.

\bibitem{wang:2021reconstructing}
Lingxiao Wang, Shuzhe Shi, and Kai Zhou.
\newblock Reconstructing spectral functions via automatic differentiation.
\newblock {\em arXiv:2111.14760}, November 2021.

\end{thebibliography}
\newpage

\end{document}